\documentclass[aps,prl, twocolumn] {revtex4}
\usepackage{epsfig}

\begin{document}
\headsep 2.5cm
\title{Disordered Carbon nanotube alloys in the Effect Medium Super Cell Approximation.}

\author{Rostam Moradian}
\email{rmoradian@razi.ac.ir}
\affiliation{$^{1}$Physics Department, Faculty of Science, Razi
University, Kermanshah, Iran\\
$^{2}$Computational Physical Science Research Laboratory, Department of Nano-Science, Institute for Studies in Theoretical Physics and Mathematics (IPM)
            ,P.O.Box 19395-1795, Tehran, Iran}

\date{\today}

\begin{abstract}
We investigate a disordered single-walled carbon nanotube  (SWCNT)  in 
an effective medium super cell approximation (EMSCA).  
 First type of disorder that we consider is the presence of vacancies. 
 Our results show 
that the vacancies induce some bound states on their neighbor host sites, leading to the creation of a band around the  Fermi energy in the SWCNT average  
density of states.Second type of disorder considered is a substitutional $B_{cb}N_{cn}C_{1-cb-cn}$ alloy due to it's applications in hetrojunctions. We found that for a fixed boron (nitrogen) concentration, by increasing the nitrogen  
(boron) concentration the averaged semiconducting gap, $E_{g}$, decreases and at a critical concentration it disappears. A consequence of our results for  nano electronic devices is that by changing the boron(nitrogen) concentration, one can make a semiconductor SWCNT with a pre-determined  energy gap. 
\end{abstract}


\maketitle

 The  role of disorder in a SWCNT is of importance from two prospectives; first in the growth process of a SWCNT due to the experimental environment some impurity atoms are inserted and vacancies are created\cite{Dresselhaus:92, Brabec:94, ebbesen:95}. Second, we deliberately implant the impurity so as to construct a new nanotube alloys, such as  $B_{cb}C_{1-cb}$, $N_{cn}C_{1-cn}$ and $B_{cb}N_{cn}C_{1-cb-cn}$ SWCNTs\cite{Miyamoto1:94, Lammert:01, Yoshioka:03}, with pre-determined  physical properties. In the first case, the effect of a point-like defect was investigated by calculation of electron reflection coefficient \cite{Kostyrko:99}, and also two substitute defects in an armchair SWCNT  \cite{Song:02}.It has been found that the symmetry of defects strongly affected the conductance and the local density of states. By different techniques, the boron nitride SWCNTs junctions \cite{Blase:97, Ferreira:00}, the spin polarization in a quasi one dimensional C/BN nanotube\cite{Choi:03} and also the current distribution in  boron and nitrogen doped SWCNTs were investigated  \cite{Liu:04}. For finite impurity concentration, a systematic field theory technique beyond single-site T-matrix approximation has not yet been applied to the disordered SWCNTs\cite{mahan:04}. In this Paper, for the first time, by applying the EMSCA\cite{Moradian1:02, Moradian2:03} method to the disordered SWCNT, we will go beyond this approximation and consider  the presence of finite impurities. We provide a more realistic description of the effects of disorder, due to  vacancies, on an armchair SWCNT's and a zigzag SWCNT's density of states (DOS). Also in this formalism, we address the question of how the doping of a  zigzag SWCNT by boron (nitrogen), i.e. $B_{cb}N_{cn}C_{1-cb-cn}$, controls the semiconducting gap, $E_{g}$.


Let us consider the Hamiltonian as a general random tight-binding model\cite{Moradian1:02}        
\begin{equation}
H=-\sum_{ij\alpha\beta\sigma}t^{\alpha\beta}_{ij}{c^{\alpha}}^{\dagger}_{i\sigma}c^{\beta}_{j\sigma}+\sum_{i\alpha\sigma} (\varepsilon^{\alpha}_{i}-\mu)
 \hat{n}^{\alpha}_{i\sigma},
\label{eq:Hamiltonian}
\end{equation}
where $t^{\alpha\beta}_{ij\sigma\sigma}$ are the hopping integrals between the $\pi$ orbitals of sites $i$ and $j$ with spin $\sigma$. $\alpha$ and $\beta$ refer to the A or B sites, $\mu$ is the chemical potential and $\varepsilon^{\alpha}_{i}$ is the random on-site energy where it takes  $0$ with probability $1-c$ for host sites and $\delta$ with probability $c$ for impurity sites. For the $B_{cb}N_{cn}C_{1-cb-cn}$ SWCNT alloy, $\varepsilon^{\alpha}_{i}$ takes $\delta$ $(=t)$ with probability $cb$ for boron sites, -$\delta$ $(=-t)$ with probability $cn$ for nitrogen sites and 0 with probability $1-cb-cn$ for the  carbon sites \cite{Yoshioka:03}, where $t$ is hopping integral to the nearest neighbour. The matrix form of Eq.\ref{eq:Hamiltonian} is, 
\begin{equation}
H=-\sum_{ij\sigma}{\Psi}^{\dagger}_{i\sigma}\hat{t}_{ij}\Psi_{j\sigma}+\sum_{i\sigma} {\Psi}^{\dagger}_{i\sigma} ({\hat\varepsilon}_{i}-\mu{\bf I}){\Psi}_{i\sigma},
\label{eq:matrixHamiltonian}
\end{equation}
where the two-component field operator, ${\Psi}^{\dagger}_{i\sigma}$, is given by
\begin{equation}
\Psi_{i\sigma}= \left(
       \begin{array}{c}
c^{A}_{i \sigma} \\c^{B}_{i \sigma} 
\end{array}\right),
\label{eq:twosite field}
\end{equation}

 and ${\hat{\varepsilon}}_{i}$ is the random on-site energy matrix,
\begin{equation}
{\hat{\varepsilon}}_{i}=\left(
       \begin{array}{cc}
\varepsilon^{A}_{i} & 0\\
0 &\varepsilon^{B}_{i}
\end{array}\right),
\label{eq:twosite random energy}
\end{equation}
 and ${\hat{t}}_{ij}$  is the hopping matrix defined by,
\begin{equation}
{\hat{t}}_{ij}=\left(
       \begin{array}{cc}
t^{AA}_{ij} & t^{AB}_{ij} \\
 t^{BA}_{ij}&t^{BB}_{ij}
\end{array}\right),
\label{eq:twosite hopping integral}
\end{equation}
 and ${\bf I}$ is a $2\times 2$ unitary matrix.

The equation of motion for electrons in such a lattice is,  
\begin{equation}
\sum_{l} \left(
       \begin{array}{c}
(E{\bf I}-{\hat\varepsilon}_{i}+{\hat\mu}_{i})\delta_{il}-{\hat t}_{il}\end{array}\right){\bf G}(l,j;E)={\bf I}\delta_{ij}
\label{eq:equation of motion}
\end{equation}
where ${\bf G}(i,j; E)$ is the  random Green function matrix defined by
\begin{equation}
 {\bf G}(i,j; E)=\left(
       \begin{array}{cc}
 G^{AA}(i,j; E) & G^{AB}(i,j; E)\\
 G^{BA}(i,j; E) & G^{BB}(i,j; E)
\end{array}\right).
\label{eq:twosite Green function}
\end{equation}

 We considered the ${\hat\varepsilon}_{i}$ as a perturbation parameter, hence ${\bf G}(i,j;E)$ in Eq.\ref{eq:equation of motion}, may be expanded in terms of the perfect Green function matrix ${\bf G}^{0}(i,j;E)$ as,
\begin{equation}
 {\bf G}(i,j;E)={ \bf G}^{0}(i,j;E)+\sum_{l}{\bf G}^{0}(i,l;E)
{\hat \varepsilon}_{l}{G}(l,j;E)
\label{eq:expanding interms of random onsite potential}
\end{equation}
where ${\bf G}^{0}(i,j;E)$ is given by

\begin{equation}
{\bf G}^{0}(i,j;E)=\frac{1}{N}\sum_{\bf k}e^{\imath{\bf k}.{\bf r}_{ij}}
\left(\begin{array}{c}
E{\bf I}-{\hat\epsilon}_{\bf k}+{\bf I}\mu 
\end{array}\right)^{-1}.
\label{eq:clean}
\end{equation}
with ${\hat \epsilon}_{\bf k}=\frac{1}{N}\sum_{ij}{\hat t}_{ij}e^{\imath{\bf k}.{\bf r}_{ij}}$ being the band structure for perfect system. In our calculations we assumed allowed  hopping to the nearest neighbors and neglected the others. Hence 
\begin{equation}
{\hat{t}}_{<ij>}=\left(
       \begin{array}{cc}
0 & t^{AB}_{<ij>} \\
 t^{BA}_{<ij>}& 0
\end{array}\right),
\label{eq:twosite hopping}
\end{equation}
and the dispersion relation is
\begin{equation}
{\hat{\epsilon}}_{\bf k}=\left(
       \begin{array}{cc}
0 & t\gamma({\bf k}) \\
 t\gamma^{*}({\bf k}) & 0
\end{array}\right).
\label{eq:dispersion relation}
\end{equation}
where $\gamma({\bf k})=\sum_{i=1}^{3}e^{\imath{\bf k}.{\bf \tau}_{i}}$ and $\tau_{i}$ are three vectors that connect an A(B) site to it's nearest neighbors B(A) sites.

The Dyson equation for the averaged Green function, $\bar{\bf G}(i,j;E)$, corresponding to Eq.\ref{eq:expanding interms of random onsite potential} is  
\begin{eqnarray}
 \bar{\bf G}(i,j;E)&=&{\bf G}^{0}(i,j;E)\nonumber \\
&+&\sum_{ll^{'}}{\bf G}^{0}(i,l;E)
{\bf\Sigma}(l,l^{'};E){\bar{\bf G}}(l^{'},j;E),
\label{eq:Dyson equation}
\end{eqnarray}
where the self energy ${\bf\Sigma}(l,l^{'};E)$ is defined by
\begin{equation}
\langle {\hat\varepsilon}_{l}{\bf G}(l,j;E)\rangle = \sum_{l^{'}}{\bf\Sigma}(l,l^{'};E){\bar{\bf G}}(l^{'},j;E).
\label{eq:self energy definition}
\end{equation}
The Fourier transform of  $\bar{\bf G}(i,j;E)$ in Eq.\ref{eq:Dyson equation} is given by 
 \begin{equation}
{ \bar {\bf G}}(i,j;E)=\frac{2}{N}\sum_{\bf k}e^{\imath{\bf k}.{\bf r}_{ij}}
\left(\begin{array}{c}
E {\bf I}-{\hat\epsilon}_{\bf k}+\mu{\bf I}-{\bf\Sigma}({\bf k};E) 
\end{array}\right)^{-1} 
\label{eq:exact average green function}
\end{equation}
where 
 \begin{equation}
{\bf\Sigma}({\bf k};E)=\frac{2}{N}\sum_{i,j}e^{-\imath{\bf k}.{\bf r}_{ij}}
{\bf\Sigma}(i,j;E),
\label{eq:self energy fourier transform}
\end{equation}
is the self energy Fourier transform.

\begin{figure}
\centerline{\epsfig{file=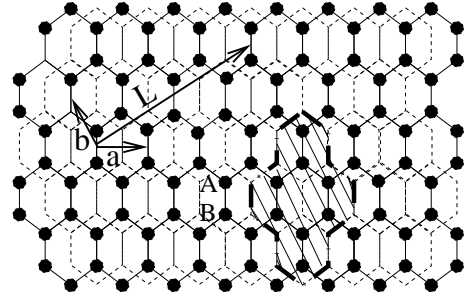  ,width=6.0cm,angle=0}}
\caption{ A two dimensional graphene sheet. The light dashed lines illustrate the Bravias lattice unit cells, ${\bf a}$ and ${\bf b}$ are the primitive vectors. Each cell includes two non-equivalent sites, which are denoted by A and B. ${L}=n_{a}{\bf a}+n_{b}{\bf b}$ is the chiral vector. For an armchair SWCNT $n_{a}=n$ and $n_{b}=2n$, while for a zigzag SWCNT $n_{a}=n$ and $n_{b}=0$. The heavy dashed line denotes a four-site super cell.
 \label{figure:1carbon1}}
\end{figure}

We solve Eq.\ref{eq:expanding interms of random onsite potential} using the  EMSCA method \cite{Moradian1:02, Moradian2:03} for the case of four sites super cell, i.e. $N_{c}=4$.
 Fig.\ref{figure:1carbon1} shows a 2-dimensional  graphene sheet. Each cell of the  Bravias lattice includes two nonequivalent sites that are denoted by A and B. The primitive vectors of the Bravias lattice are ${\bf a}$ and ${\bf b}$ and the chiral vector is ${\bf L}$. The heavy dashed line on the figure shows a four-sites super cell of the graphene Bravias lattice.  

 In the EMSCA technique, the super cell random Green functions, ${\bf G}^{im}_{sc}(i,j;E)$,  are related to the cavity Green function $\hat{\mathcal G}(i,j;E)$ via
\begin{equation}
 {\bf G}^{im}_{sc}(I,J;E)=\hat{\mathcal G}(I,J;E)+\sum_{L}\hat{\mathcal G}(I,L;E)
{\hat \varepsilon}_{L}{\bf G}^{im}_{sc}(L,J;E),
\label{eq:random super cell Green function}
\end{equation}
where $\{I\}$ refers to the sites inside the super cell. Also the Dyson's-like equation for the average super cell Green function, ${\bar{\bf G}}_{sc}(I,J;E)$, is given by
\begin{eqnarray}
 {\bar{\bf G}}_{sc}(I,J;E)&=&\hat{\mathcal G}(I,J;E)\nonumber\\&+&\sum_{LL^{'}}\hat{\mathcal G}(I,L;E)
{\bf \Sigma}_{sc}(L,L^{'};E){\bf G}(L^{'},J;E).\nonumber\\
\label{eq:Dyson like equation}
\end{eqnarray}
The Fourier transform of  ${\bar{\bf G}}_{sc}(I,J;E)$ in Eq.\ref{eq:Dyson like equation} is 
\begin{equation}
 {\bar{\bf G}}_{sc}({\bf K}_{n};E)=\hat{\mathcal G}({\bf K}_{n};E)+\hat{\mathcal G}({\bf K}_{n};E)
{\bf \Sigma}_{sc}({\bf K}_{n};E){\bf G}({\bf K}_{n};E)
\label{eq:fourier transform of Dyson like equation},
\end{equation}
where 
\begin{equation}
\Sigma({\bf K}_{n};E)=\frac{1}{Nc}\sum_{IJ} e^{{\bf K}_{n}.{\bf r}_{IJ}}\Sigma(I,J;E) 
\label{eq:k-equality of self energies}
\end{equation}
and
\begin{eqnarray}
{\bar G}({\bf K}_{n};E)&=&
\frac{N_{c}}{N}\nonumber\\&\times&\sum_{{\bf k}\in\; nth\; patches}\left(\begin{array}{c}
{\bf I}E-{\hat\epsilon}_{\bf k}+{\bf I}\mu-{\bf \Sigma}({\bf K}_{n};E) 
\end{array}\right)^{-1}.\nonumber\\ 
\label{eq:k-supercell average green function}
\end{eqnarray}
To calculate the ${\bar G}_{sc}(I,J;E)$ and $G^{imp}_{sc}(I,J;E)$, Eqs.\ref{eq:random super cell Green function}-\ref{eq:k-supercell average green function} should be solved self consistently.

\begin{figure}
\centerline{\epsfig{file=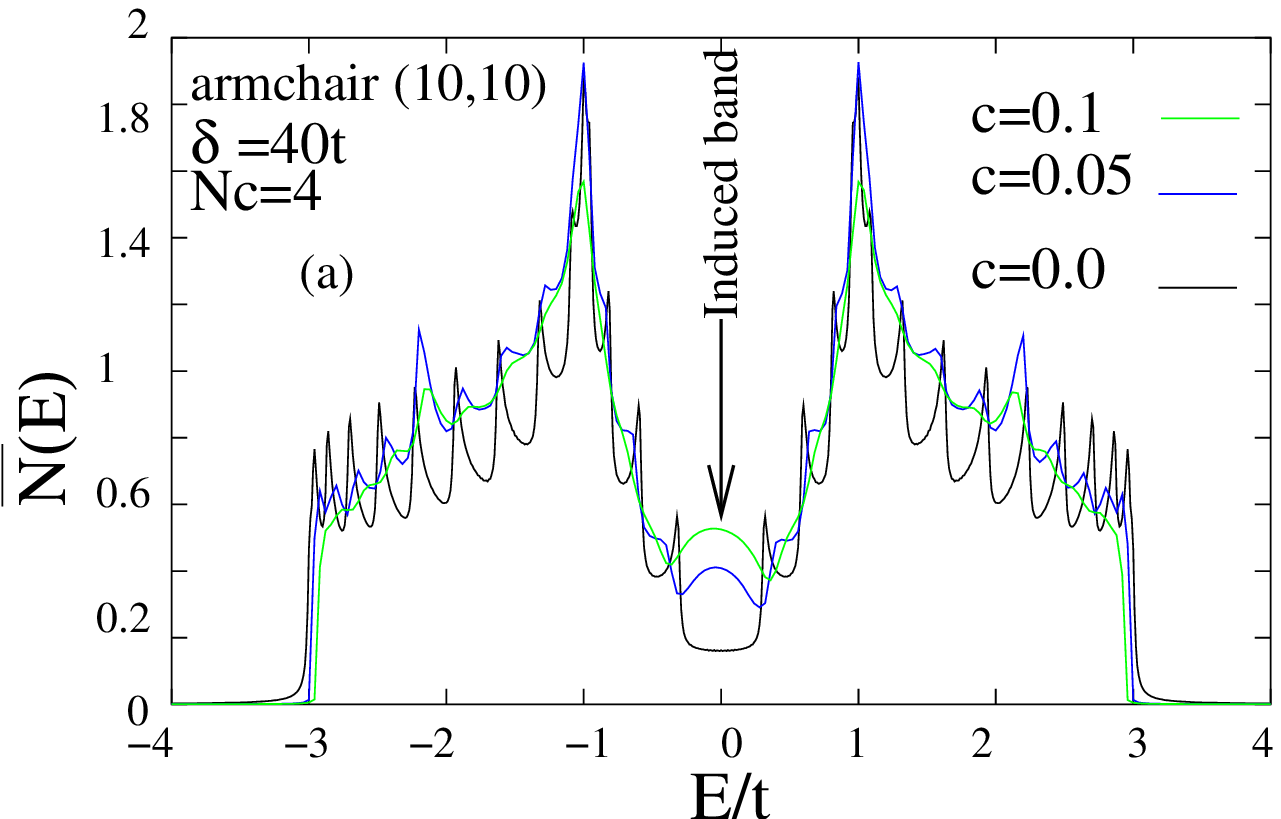  ,width=6.0cm,angle=0}}
\centerline{\epsfig{file=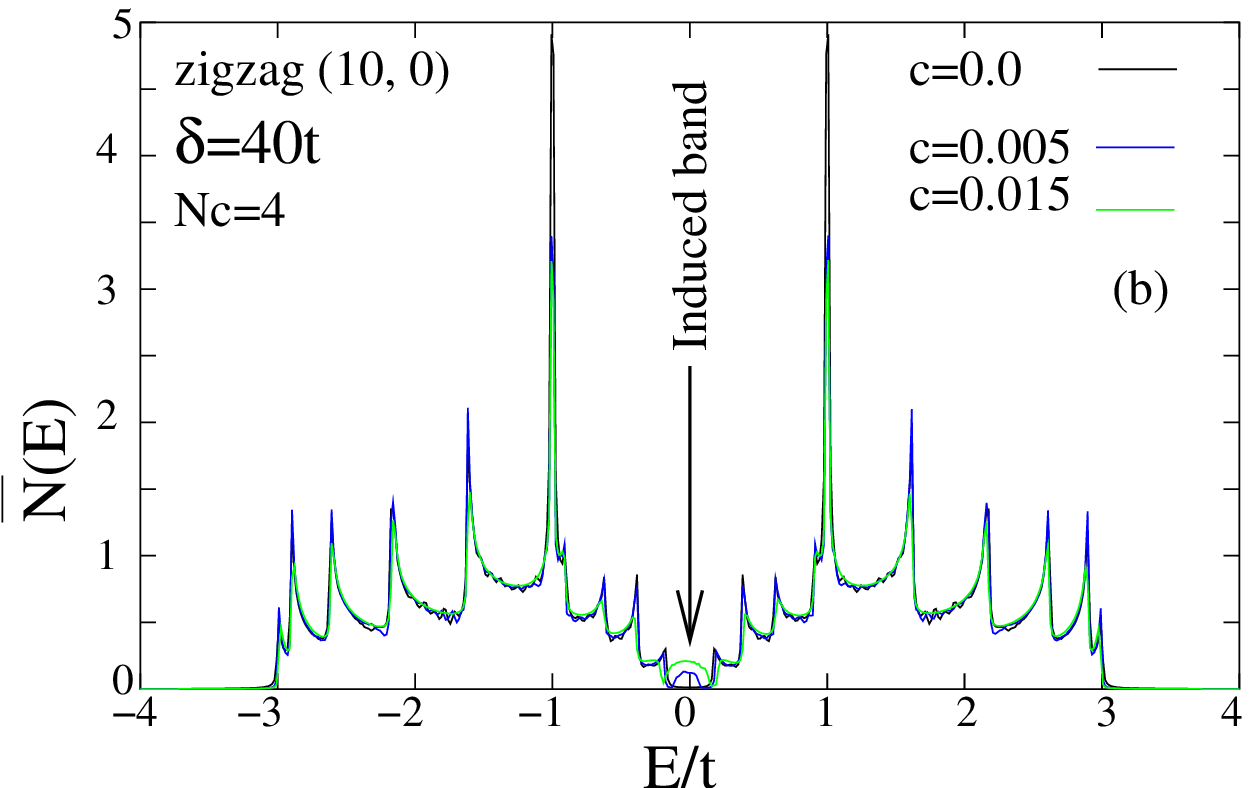 ,width=6.0cm,angle=0}}
\caption{(a) Comparison of average density of states for a: $(10,10)$ armchair SWCNT and (b): for a $(10,0)$ zigzag SWCNT at $\delta=40t$ and different impurity concentration. At high random energies, $\delta>> band \;width$, we have band splitting and the impurity band is located at higher energies, while remaining sites are the vacancies that induce a band around the Fermi energy.  By increasing the vacancy concentration the height of the peak at the Fermi energy increases and the van How singularities are smeared out.
 \label{figure:4carbon1} }
\end{figure}

 A SWCNT with vacancies is considered, the averaged density of states for different vacancy concentrations is calculated. We found that vacancies create some bound states around the Fermi level on their host neighbour sites, hence constructing a band in the averaged density of states. Also the one-dimensional (1D) van How singularities in high vacancy concentrations disappear. Fig.\ref{figure:4carbon1}(a),(b) shows the comparison between the average density of states for different vacancy concentrations in $(10,10)$ and  $(10,0)$ SWCNTs respectively. The bound states due to vacancies around the Fermi energy is  marked by an arrow. In short, our results show that vacancies not only change the average density of states but also the number of electrons located on the host sites and also at high vacancy concentrations SWCNT's loses their 1D characteristics and become similar to a 2D disordered geraphene sheet. 
 
\begin{figure}
\centerline{\epsfig{file=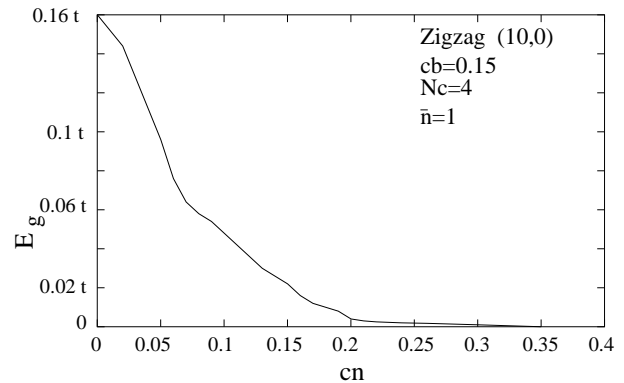 ,width=8.0cm,angle=0}}
\caption{The figure shows the effects of nitrogen doping on semiconducting gap, $E_{g}$, for a $(10,0)$ zigzag SWCNT alloy. 
 \label{figure:5carbon1}}
\end{figure}

\begin{figure}
\centerline{\epsfig{file=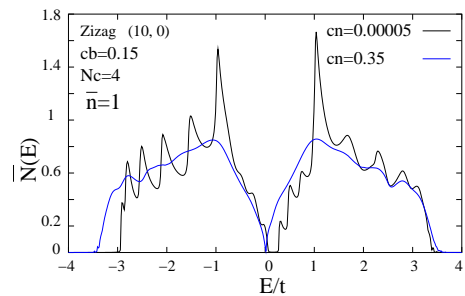 ,width=6.0cm,angle=0}}
\caption{
The Effects of nitrogen doping on a $(10,0)$ zigzag SWCNT,s average density of states. At a fixed boron concentration $cb=0.15$, the average density of states for two nitrogen concentrations, low $cn=0.0005$ and critical $cn=0.35$, are compared. At the critical concentration $E_{g}$ is zero and the van How singularities  disappear. 
 \label{figure:6carbon1}}
\end{figure}

We now investigate the effect of nitrogen and boron doping on a $(10,0)$ zigzag SWCNT. Two cases are considered, first fixed boron concentration at $cb=0.15$, with variable nitrogen concentration. In this case, we found that the average semiconducting gap, $E_{g}$, decreased by increasing the nitrogen concentration, and at a critical concentration of $cn=0.35$ it disappeared. Fig.\ref{figure:5carbon1} illustrates the effects of the nitrogen doping on the $E_{g}$. 
To clarify our results, we compare the average density of states for low and critical nitrogen concentrations. Fig.\ref{figure:6carbon1} compares the average density of states for the low, $cn=0.00005$, and critical, $cn=0.35$, nitrogen concentration. At $cn=0.00005$, the gap is  located at the top of the Fermi energy, but inside the pure SWCNT conduction band. By increasing the nitrogen concentration, the low edge of the conduction band is moved until the gap is closed, hence a semiconductor to semi-metal phase transition takes place. Also 1D van How singularities disappeared.   

In the second case, we fixed the nitrogen concentration at $cn=0.1$, while varying the boron concentration. We found that $E_{g}$, decreases with an increase in the boron concentration, and at a critical concentration; $cb=0.35005$ it tended to zero.  
\begin{figure}
\centerline{\epsfig{file=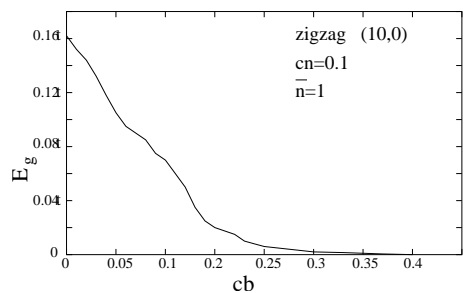,width=6.0cm,angle=0}}
\caption{Figure shows the effects of boron doping on the semiconducting gap, $E_{g}$, in a $(10,0)$ zigzag SWCNT alloy. 
 \label{figure:7carbon1}}
\end{figure}
Fig.\ref{figure:7carbon1} compares the average density of states for the low, $cb=0.00005$, and critical, $cb=0.35005$, boron concentrations. For this case, the $E_{g}$ is closed, similar to the first case, and the van How singularities also disappeared. Furthermore, the semiconductor to semi-metal phase transition was also observed.   
\begin{figure}
\centerline{\epsfig{file=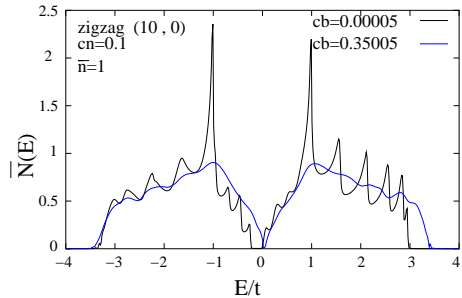,width=6.0cm,angle=0}}
\caption{Effects of boron doping on a $(10,0)$ zigzag SWCNT's average density of states. At a fixed nitrogen concentration of $cn=0.1$, the average density of states for two boron concentrations $cb=0.0005$ and $cb=0.35005$ are compared. At a critical concentration, $E_{g}$ is closed and the van How singularities disappears and a semiconductor semi-metal phase transition takes place.  
 \label{figure:8carbon1}}
\end{figure}

In conclusion, we have applied the EMSCA method to a disordered SWCNT in order  to investigate and role  of disorder in such materials.  For a $(10, 10) $ armchair tube and also a zigzag $(10, 0)$ tube we found that the vacancies induce some bound states on their host neighbor sites, creating a band around the Fermi energy in the average density of states. The consequences of this band formation around the Fermi energy and also disappearance of the 1D van How singularities at high vacancy concentrations is; that the density of states of an armchair and also a zigzag SWCNT become similar to a disordered (vacancy disorder) 2D graphene sheet density of states.
A $(10,0)$ zigzag $B_{cb}N_{cn}C_{1-cb-cn}$  SWCNT alloy was investigated. We 
found  that for a fixed boron (nitrogen) concentration, by increasing the nitrogen (boron) concentrations, the $E_{g}$ decreases and at a critical concentration it becomes closed. Therefore, a semiconductor to a semi-metal phase transition takes place. Our results show that we can control the $E_{g}$ by changing the nitrogen (boron) concentration. 

\acknowledgements I would like to thanks professor Rafii-Tabar for helpful discussion.


\end{document}